\documentclass[preprint,aip,apl]{revtex4-1}
\usepackage{subfigure}
\usepackage{graphicx}

\begin{document}

%Title page:

\title[Controlled-NOT gate operating with single photons]{Controlled-NOT gate operating with single photons}

\author{M.~A.~Pooley}
 \email{map57@cam.ac.uk}
 \affiliation{Toshiba Research Europe Limited, Cambridge Research Laboratory, 208~Science~Park, Milton Road, Cambridge, CB4 0GZ, U.K.}
 \affiliation{Cavendish Laboratory, Cambridge University, J.~J.~Thomson Avenue, Cambridge, CB3 0HE, U.K.}
\author{D.~J.~P.~Ellis}
 \affiliation{Toshiba Research Europe Limited, Cambridge Research Laboratory, 208~Science~Park, Milton Road, Cambridge, CB4 0GZ, U.K.}
\author{R.~B.~Patel}
 \affiliation{Toshiba Research Europe Limited, Cambridge Research Laboratory, 208~Science~Park, Milton Road, Cambridge, CB4 0GZ, U.K.}
 \affiliation{Cavendish Laboratory, Cambridge University, J.~J.~Thomson Avenue, Cambridge, CB3 0HE, U.K.}
\author{A.~J.~Bennett}
 \affiliation{Toshiba Research Europe Limited, Cambridge Research Laboratory, 208~Science~Park, Milton Road, Cambridge, CB4 0GZ, U.K.}
\author{K.~H.~A.~Chan}
 \affiliation{Toshiba Research Europe Limited, Cambridge Research Laboratory, 208~Science~Park, Milton Road, Cambridge, CB4 0GZ, U.K.}
 \affiliation{Cavendish Laboratory, Cambridge University, J.~J.~Thomson Avenue, Cambridge, CB3 0HE, U.K.}
\author{I.~Farrer}
 \affiliation{Cavendish Laboratory, Cambridge University, J.~J.~Thomson Avenue, Cambridge, CB3 0HE, U.K.}
\author{D.~A.~Ritchie}
 \affiliation{Cavendish Laboratory, Cambridge University, J.~J.~Thomson Avenue, Cambridge, CB3 0HE, U.K.}
\author{A.~J.~Shields}
 \affiliation{Toshiba Research Europe Limited, Cambridge Research Laboratory, 208~Science~Park, Milton Road, Cambridge, CB4 0GZ, U.K.}

\begin{abstract}

The initial proposal for scalable optical quantum computing required
single photon sources, linear optical elements such as beamsplitters
and phaseshifters, and photon detection. Here we demonstrate a two
qubit gate using indistinguishable photons from a quantum dot in a
pillar microcavity. As the emitter, the optical circuitry,
and the detectors are all semiconductor, this is a promising approach
towards creating a fully integrated device for scalable quantum computing.

\end{abstract}

\maketitle

%%%%%%%%%%%%%%%%%%%%%%%%%%%%%%%%%%%%%%%%%%%%%%%%%%%%%%%%%%%%%%%%%%%%%%%%%%%%%%%%%%%%%%%%%%%%%%%%%%

%%%%%%%%%%%%%%%%%%%%%%%%%%%%%%%%%%%%%%%%%%%%%%%%%%%%%%%%%%%%%%%%%%%%%%%%%%%%%%%%%%%%%%%%%%%%%%%%%%
% Main text of the letter

Quantum computing offers a drastic increase in the speed of solving
parallelizable computational tasks. Optical quantum computing, where
photons are used as quantum bits (qubits), has many advantages. Photons
have long coherence times and travel fast, allowing connection between
remote nodes of a quantum computer, and they can easily be manipulated at the
single qubit level. A promising approach to optical quantum computing,
linear optics quantum computing (LOQC)\cite{Knill2001}, requires
triggered single photon sources which supply photons into optical
modes in well-defined states. Several protocols have been proposed
that reduce the resource overhead required to build an
optical quantum computer \cite{Raussendorf2001,Browne2005,Kok2007},
but the ability to initialize a certain photon state is still of
central importance. Significant progress in experimental work on
few-qubit gates has been made using probabilistic sources, such as
optically pumped parametric down-conversion crystals
\cite{Gasparoni2004,O'Brien2003,Okamoto2005,Politi2008}. However, the
Poissonian statistics which govern such sources can seriously impair the
performance of quantum logic gates, precluding systems containing many gates.

This has motivated much research into on-demand single photon emission.
A variety of triggered single photon sources have been demonstrated
including, molecules\cite{Lounis2000,Brunel1999}, trapped
ions\cite{Kuhn2002}, color centers, and quantum dots
(QDs)\cite{Michler}. Semiconductor sources are particularly promising as
many emitters can be fabricated on a small area and they can
be integrated directly into semiconductor waveguide
technology. Here we present a two qubit
gate in which a single QD is used as the photon source and the
optical circuitry is realized using a semiconductor waveguide.
The Controlled-NOT (CNOT) gate we demonstrate is the basic building block
of quantum logic, since in combination with one qubit gates it can be used
to perform any quantum operation.

\begin{figure}%
\includegraphics[width=80mm]{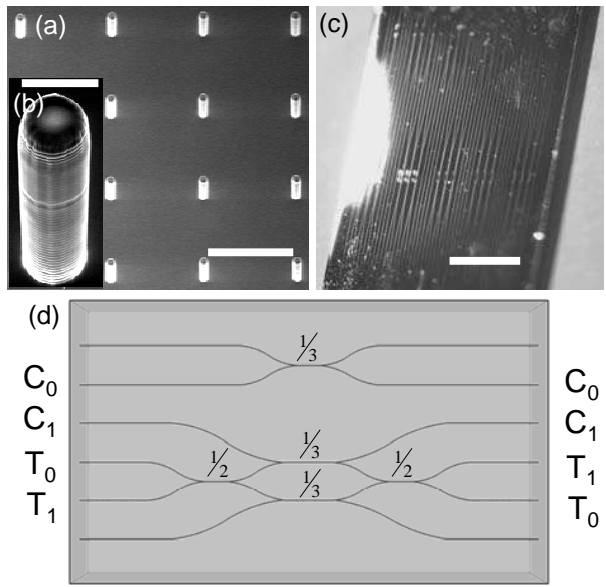}%
\caption{(Color online) SEM images of (a) an array of pillar microcavities, and (b) a single pillar microcavity
 (scale bars 20 $\mu$m and 2 $\mu$m respectively). (c) a semiconductor waveguide chip (scale bar 5 mm)
 (d) Schematic diagram of a single waveguide
circuit.}%
\label{fig:1}%
\end{figure}

In our work, the photon source consists of a single self-assembled InAs QD embedded inside a 1.5 $\mu m$ diameter pillar microcavity with a quality factor, $Q\sim{9000}$. The QD exciton transition emits at a wavelength of $\lambda=931$ nm, which is resonant with the HE$_{11}$ cavity mode of the pillar. The planar structure of the pillar microcavity, which is grown by molecular beam epitaxy, consists of 17 (25) GaAs/AlGaAs distributed Bragg reflector periods above (below) a one-wavelength thick GaAs cavity centered on a single layer of self-assembled InAs QDs. The pillar is defined using reactive ion etching, see Figures 1a - b.  The advantage of using a
pillar microcavity structure is two-fold, (1) Purcell
enhancement\cite{Purcell1946} reduces the radiative lifetime
relative to the coherence time, increasing the probability that
consecutive photons are indistinguishable; (2) Collection efficiency
is significantly increased as QD emission is coupled into the cavity
mode.

The pillar is held at a temperature of 4.5K and the QD is excited quasi-resonantly one LO-phonon ($\sim{32}$ meV) above the exciton energy level by a pulsed 1364 meV modelocked laser with a repetition frequency of 80 MHz. Quasi-resonant excitation prevents excess carrier generation in the wetting layer, thus reducing the effective radiative lifetime and increasing the coherence time of the resulting excitons. In addition, excitation through a phonon resonance allows selective excitation of only one polarization state, as the polarization of the laser photon is transferred to the spin of the exciton.  This doubles the internal quantum efficiency of the source, relative to quasi-resonant excitation schemes that promote the system into an excited state which can decay into either of the exciton polarization states. Exciting the source under these conditions, at a small fraction of saturation, we measure a radiative lifetime of $\tau_R=106$ ps, a coherence time of $\tau_C=148$ ps, and the photon emission rate is $\sim{5}$ MHz into an NA$=0.42$ microscope objective above the emitter.

The CNOT input state requires two photons, both of which are emitted from the same QD. An unbalanced Michelson interferometer is used to split each laser pulse into two pulses separated by 1.95 ns, thus allowing the QD to be excited twice in quick succession each laser repetition cycle, as required for our state preparation scheme.

\begin{figure}%
\includegraphics[width=80mm]{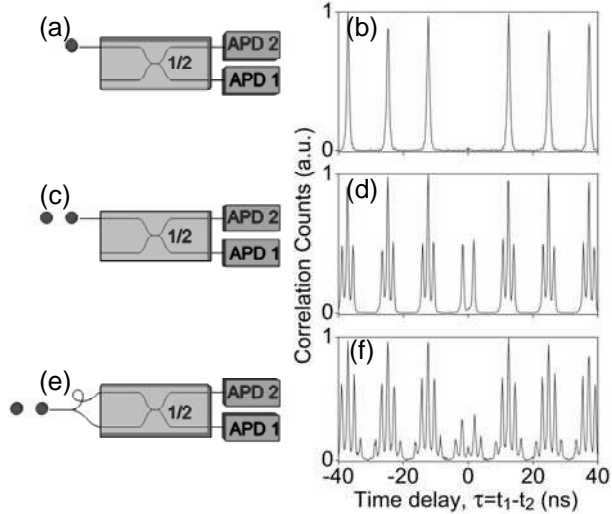}%
\caption{(Color online) (a,b), Pulsed second-order autocorrelation
measurement yields $g^{(2)}(0)=0.013\pm0.004$. (c,d), Pulsed
autocorrelation measurement when the QD is excited twice in quick
succession shows that the probability of either excitation resulting
in multi-photon emission to be $g=0.0063\pm0.0005$. (e,f),
Pulsed two photon interference measurement using a $\frac{1}{2}$
coupler yields a two photon interference visibility of
$V^{(2)}=0.72 \pm 0.05$.}%
\label{fig:2}%
\end{figure}

The single
photon nature of the source is confirmed by a pulsed second-order
autocorrelation measurement in a waveguide circuit shown in Figure 2a.
The results, seen in Figure
2b, show that the multi-photon emission rate is just $(1.3 \pm
0.4)\%$ that of a Possonian emitter of the same intensity. Also
shown in Figure 2d is a second-order autocorrelation measurement
when the QD is excited twice per cycle,
under the conditions needed to implement the CNOT operation. From
this measurement it is possible to extract the probability that
either of the emission events are multi-photon, $g$. The residual
between the data and a double Lorentzian fit to the peaks at $\pm$
1.95 ns yields a probability of multi-photon emission in either
pulse of $g=0.0063 \pm 0.0005$.

Each pair of photons must be indistinguishable in order to maximize
the degree of two photon interference, $V^{(2)}$, which determines the
probability of the two qubits interacting correctly when the input state has
$|1\rangle_C$. A two photon interference measurement\cite{Hong1987}, shown
in Figure 2f, using an isolated $\frac{1}{2}$ directional coupler,
 yields $V^{(2)}=0.72\pm0.05$. This is comparable with the highest
reported for single photon semiconductor sources\cite{Santori2002,Bennett2008}.

The optical circuit is realized using a semiconductor waveguide,
shown in Figure 1c, in which directional couplers are used to
perform the function of beamsplitters. The circuit implements the
CNOT gate proposed by Ralph \textit{et al.}\cite{Ralph2002}
and consists of a network of directional couplers with coupling ratios of
$\frac{1}{2}$ and $\frac{1}{3}$ as shown schematically in Figure 1d.
Each waveguide core has cross-sectional dimensions of $3.8\times3.8$ $\mu$m,
and the complete CNOT circuit has lateral dimensions of $1.25\times32.5$ mm.
The semiconductor waveguide is fabricated in silica on a silicon
substrate using standard semiconductor processing. Photons are
confined within the waveguides by the refractive index contrast
between the core and cladding regions. The coupling ratio of a
directional coupler is wavelength dependent and can be precisely
tuned by adjusting the interaction length. Isolated test couplers,
nominally identical to those in the network, were formed adjacent
to the circuit to permit the direct measurement of the actual
coupling ratios inside the CNOT circuit. The couplers used in this
work have coupling ratios of $R_{1/3}=0.345$ and $R_{1/2}=0.495$.

The two qubits are path encoded, considering the logical basis where
each qubit has a single photon in one path and zero photons in the
other, the input states are $|C\rangle |T\rangle = |CT\rangle =$
$|00\rangle_{CT},|01\rangle_{CT},|10\rangle_{CT},|11\rangle_{CT}$.
The QD emission is coupled into a 50/50 fiber beamsplitter, with one output
connected to a target path input and the other output connected to a
control path input via a 1.95 ns delay. If the first photon takes
the control path and the second takes the target path the two
photons enter the circuit simultaneously and the desired input state
is realized. Light is coupled into and out of the waveguide using fiber
arrays which are butt-coupled with index matching fluid to the waveguide.
Polarization-maintaining (PM) single-mode fiber is used for the inputs and
multi-mode fiber is used to collect from the outputs. Optical loss in
the circuit is negligible, but typically the insertion loss coupling from
PM fibre to a waveguide is 50\%.

\begin{figure}%
\includegraphics[width=80mm]{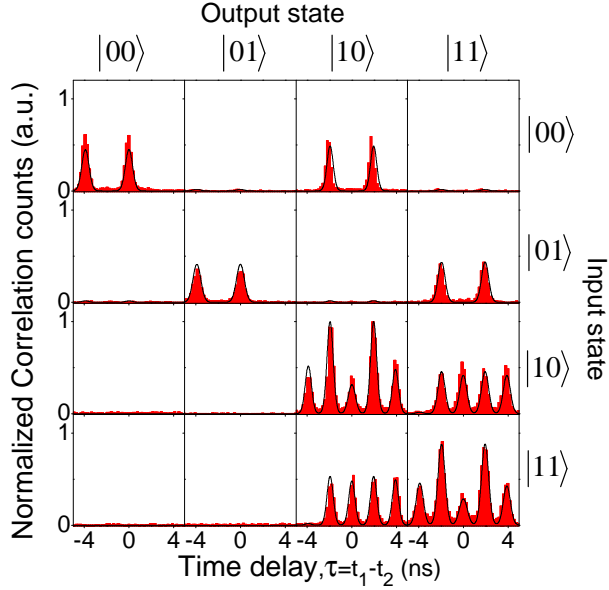}%
\caption{(Color online) Each pane corresponds to a correlation measurement for a
single input and output combination. Measured data are shown as
filled red bars. Solid black lines show calculated curves for the
expected correlations given the measured coupler ratios, single
photon interference visibility, and two photon
interference visibility. The peaks at $\tau=0$
correspond to events where both photons entered the circuit
simultaneously, and the area of these peaks is used to extract the
truth table for the operation.}%
\label{fig:3}%
\end{figure}

For each input state, the QD emission is coupled into the corresponding
waveguide inputs and four time-resolved correlation measurements are
simultaneously acquired between control and target output paths, one for
each of the possible output states. Each set of correlation measurements
requires 30 minutes accumulation time.  The coincidence counts at zero
delay ($\tau=0$) correspond to cases where the correct input state
is realized, the area of these peaks are used to construct the truth table.

Figure 3 shows the experimental correlation measurements (filled red bars),
along with predicted correlation curves (solid black lines).
The predicted curves are calculated by considering the possible paths
that the photons can take through the waveguide network, with the
probability of each path adjusted to account for the experimentally determined
coupler reflectivities and single-photon interference visibility.
For the $\langle10|10\rangle$ and $\langle11|11\rangle$ correlations the central $\tau=0$ peaks are dependent on
two-photon interference effects, these peaks are calculated using a wavepacket
approach similar to Legero \textit{et al.}\cite{Legero2004,Patel2010}, in which the temporal properties of the photons ($\tau_R$ and $\tau_C$) and the waveguide imperfections are considered.

From the
$\langle00|00\rangle$ and $\langle00|01\rangle$ correlations it is
possible to extract a single photon interference visibility of
$V^{(1)}=0.97$, which gives the probability of a single photon
traversing the Mach-Zehnder interferometer correctly in the absence
of any qubit-qubit interaction. The two photon interference
visibility, which can be obtained from the $\langle10|10\rangle$ and
$\langle11|11\rangle$ correlations, is limited not only by the
distinguishability of the photons but also by properties of
waveguide circuit, specifically the single photon interference
visibility and the non-optimal coupler ratios. The data agrees well
with the predicted correlation curves which are
calculated using the measured values of $V^{(1)}$, $R_{1/3}$, $R_{1/2}$ and $V^{(2)}=0.67$,
which is consistent with the two photon visibility obtained for an isolated $\frac{1}{2}$
coupler.

\begin{figure}%
\includegraphics[width=80mm]{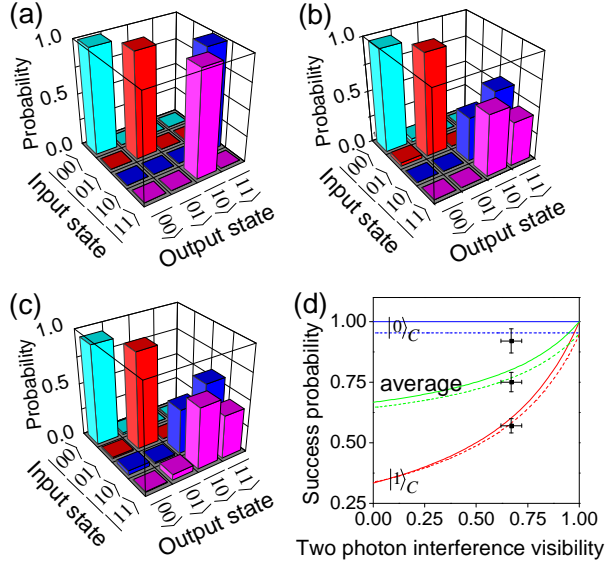}%
\caption{(Color online) (a) Truth table for an ideal CNOT gate.
(b) Predicted truth table for the circuit and photon source
used in this experiment based on the experimentally achieved two
photon interference visibility, single photon interference
visibility, and coupler ratios. (c) Experimentally achieved
truth table. (d) Success probability as a function of two
photon interference visibility for the $|0\rangle_C$ input states
(blue), $|1\rangle_C$ input states (red) and averaged over all
input states (green). Solid lines are for an ideal circuit, dashed lines are for the circuit realized experimentally. The values
extracted from the measured truth table are shown as black squares,
error bars are Poissonian errors in the number of counts.}%
\label{fig:4}%
\end{figure}

Figures 4a-c show the measured truth table, along with those for
ideal CNOT gate operation, and operation with photon
interference visibilities of $V^{(1)}=0.97$ and $V^{(2)}=0.67$.
Each element of the truth tables in figure \ref{fig:4} is extracted from the area
of the coincidence ($\tau=0$) peak in the corresponding correlation measurement of figure \ref{fig:3}.
Each truth table row corresponds to a single input state, and is normalized by
the total area of the four coincidence peaks in the associated outputs. The success probability for an
input state is then given by the value of the truth table element which
corresponds to the correct output state.

The success probability as a function of $V^{(2)}$, for an ideal
CNOT circuit (solid lines) and the circuit realized experimentally in this work (dashed lines), is shown in Figure 4d. The success probabilities obtained experimentally are shown as filled black squares in the same figure. There is good agreement between the measured and calculated values. The truth table for the operation is extracted from the coincidence
measurements by integrating all counts in a window of 1.95 ns
centered at $\tau=0$. The probability of the gate succeeding
averaged over all inputs is 71\%. Taking a window of 0.6 ns
increases the overall success probability to 75\%, while preserving
60\% of the coincidences. In principle, detectors with a faster
response time could further improve the measured success rate.
It is clear that increasing the indistinguishability between photons is the key
to improving the success probability of the gate.

In our experiment the degree of photon indistinguishability is
determined by the ratio of the radiative decay time and coherence
time of the exciton within the QD. For the quasi-resonant excitation
scheme we use, residual decoherence is
caused by interaction of the exciton with the solid-state
environment. There is potential to increase the coherence time
through the use of coherent Raman excitation
schemes\cite{Santori2009,Kiraz2004}. Even better,
resonance fluorescence measurements\cite{Nguyen} can transfer the coherence properties of the excitation
laser onto the exciton photon, significantly lengthening the
coherence times toward that of the driving laser. Ideally, this
excitation scheme could be used to drive multiple quantum dots to
generate identical photons. The natural inhomogeneity in linewidth
for each dot would be suppressed if they were driven with the same
laser, and the inhomogeneity in emission energy could be suppressed
with electric field tuning of the transitions, as has
recently been demonstrated \cite{Patel2010}.

In conclusion, we have demonstrated logical operation of an all
semiconductor optical CNOT gate operating with indistinguishable single photons.
These results are promising for the
scalability of future photonic quantum computing technology. Extending our
scheme to incorporate multiple single photon sources will allow deterministic
preparation of multi-photon input states. Integration of the photon source
and detectors directly into photonic circuitry will allow further miniaturization and
reduction of the resources required.

%%%%%%%%%%%%%%%%%%%%%%%%%%%%%%%%%%%%%%%%%%%%%%%%%%%%%%%%%%%%%%%%%%%%%%%%%%%%%%%%%%%%%%%%%%%%%%%%%%%


\begin{thebibliography}{10}

\bibitem{Knill2001}
E.~Knill, R.~Laflamme, and G.~J. Milburn, {\em Nature}, vol.~409, pp.~46--52, Jan.
  2001.

\bibitem{Raussendorf2001}
R.~Raussendorf and H.~J. Briegel, {\em Phys.
  Rev. Lett.}, vol.~86, pp.~5188--5191, May 2001.

\bibitem{Browne2005}
D.~E. Browne and T.~Rudolph, {\em Phys. Rev. Lett.}, vol.~95, pp.~010501--, June 2005.

\bibitem{Kok2007}
P.~Kok, W.~J. Munro, K.~Nemoto, T.~C. Ralph, J.~P. Dowling, and G.~J. Milburn,{\em Rev. Mod.
  Phys.}, vol.~79, pp.~135--174, Jan. 2007.

\bibitem{Gasparoni2004}
S.~Gasparoni, J.-W. Pan, P.~Walther, T.~Rudolph, and A.~Zeilinger, {\em Phys. Rev. Lett.}, vol.~93, pp.~020504--, July 2004.

\bibitem{O'Brien2003}
J.~L. O'Brien, G.~J. Pryde, A.~G. White, T.~C. Ralph, and D.~Branning, {\em
  Nature}, vol.~426, pp.~264--267, Nov. 2003.

\bibitem{Okamoto2005}
R.~Okamoto, H.~F. Hofmann, S.~Takeuchi, and K.~Sasaki, {\em Phys.
  Rev. Lett.}, vol.~95, pp.~210506--, Nov. 2005.

\bibitem{Politi2008}
A.~Politi, M.~J. Cryan, J.~G. Rarity, S.~Yu, and J.~L. O'Brien,
  {\em Science}, vol.~320,
  no.~5876, pp.~646--649, 2008.

\bibitem{Lounis2000}
B.~Lounis and W.~E. Moerner, {\em Nature}, vol.~407, pp.~491--493, Sept. 2000.

\bibitem{Brunel1999}
C.~Brunel, B.~Lounis, P.~Tamarat, and M.~Orrit, {\em Phys. Rev.
  Lett.}, vol.~83, pp.~2722--2725, Oct. 1999.

\bibitem{Kuhn2002}
A.~Kuhn, M.~Hennrich, and G.~Rempe, {\em Phys. Rev. Lett.}, vol.~89,
  pp.~067901--, July 2002.

\bibitem{Michler}
P.~Michler, A.~Kiraz, C.~Becher, W.~V.~Schoenfeld, Lidong Zhang, E.~Hu, and A.~Imamoglu {\em Science},
  vol.~290, pp.~2282--2285, Dec. 2000.

\bibitem{Purcell1946}
E.~M. Purcell, {\em Phys. Rev.}, vol.~69, p.~681, 1946.

\bibitem{Hong1987}
C.~K. Hong, Z.~Y. Ou, and L.~Mandel, {\em Phys. Rev. Lett.},
  vol.~59, pp.~2044--2046, Nov. 1987.

\bibitem{Santori2002}
C.~Santori, D.~Fattal, J.~Vuckovic, G.~S.~Solomon, and Y.~Yamamoto, {\em Nature}, vol.~419, pp.~594--597, Oct. 2002.

\bibitem{Bennett2008}
A.~J.~Bennett, R.~B.~Patel, A.~J.~Shields, K.~Cooper, P.~Atkinson, C.~A.~Nicoll, and D.~A.~Ritchie, {\em Appl. Phys. Lett.}, vol.~92, pp.~193503--3, May 2008.

\bibitem{Ralph2002}
T.~C. Ralph, N.~K. Langford, T.~B. Bell, and A.~G. White, {\em Phys. Rev. A}, vol.~65,
  pp.~062324--, June 2002.

\bibitem{Legero2004}
T.~Legero, T.~Wilk, A.~Kuhn, and G.~Rempe, {\em Appl. Phys. B}, vol.~77, pp.~979--802, 2004.

\bibitem{Patel2010}
R.~B.~Patel, A.~J.~Bennett, I.~Farrer, C.~A.~Nicoll, D.~A.~Ritchie, and A.~J.~Shields, {\em Nat Photon}, vol.~4, .~632--635, 2010.

\bibitem{Santori2009}
C.~Santori, D.~Fattal, K.-M.~C. Fu, P.~E. Barclay, and R.~G. Beausoleil, {\em New Journal of Physics},
  vol.~11, no.~12, pp.~123009--, 2009.

\bibitem{Kiraz2004}
A.~Kiraz, M.~Atature, and A.~Imamoglu,
  {\em Phys. Rev. A}, vol.~69, pp.~032305--, Mar. 2004.

\bibitem{Nguyen}
H.~S.~Nguyen, G.~Sallen, C.~Voisin, Ph.~Roussignol, C.~Diederichs, and G. Cassabois,  {\em Appl. Phys. Lett.}, vol.~99, pp.~261904, Dec. 2011.



\end{thebibliography}
\end{document}